# The age structure of the Milky Way's halo

**Authors:** D. Carollo[1,2], T. C. Beers[1], V. M. Placco[1], R. M. Santucci[3], P. Denissenkov[4], P. B. Tissera[5], G. Lentner[1], S. Rossi[3], Y. S. Lee[6], J. Tumlinson[7]

[1]Department of Physics and JINA Center for the Evolution of the Elements, University of Notre Dame, 225 Nieuwland Science Hall, Notre Dame, IN 46556, USA

[2]INAF-Osservatorio Astronomico di Torino, 10025 Pino Torinese, Italy

[3]Departamento de Astronomia-Instituto de Astronomia, Geofísica e Ciências Atmosféricas, Universidade de São Paulo, São Paulo, SP 05508-900, Brazil

[4]Department of Physics & Astronomy, University of Victoria, Victoria, BC, V8W3P6, Canada

[5]Departamento de Ciencias Fisicas and Millennium Institute of Astrophysics, Universidad Andres Bello, Av. Republica 220, Santiago, Chile

[6]Chungnam National University, Daejeon 34134, Korea

[7]Space Telescope Science Institute, Baltimore, MD 21218, USA

**\*Corresponding author.** Email: dcaroll1@nd.edu

**We present a new, high-resolution chronographic (age) map of the Milky Way's halo, based on the inferred ages of ~130,000 field blue horizontal-branch (BHB) stars with photometry from the Sloan Digital Sky Survey. Our map exhibits a strong central concentration of BHB stars with ages greater than 12 Gyr, extending up to ~15 kpc from the Galactic center (reaching close to the solar vicinity), and a decrease in the mean ages of field stars with distance by 1-1.5 Gyr out to ~45-50 kpc, along with an apparent increase of the dispersion of stellar ages, and numerous known (and previously unknown) resolved over-densities and debris streams, including the Sagittarius Stream. These results agree with expectations from modern $\Lambda$CDM cosmological simulations, and support the existence of a dual (inner/outer) halo system, punctuated by the presence of over-densities and debris streams that have not yet completely phase-space mixed.**

**One Sentence Summary:** We detect numerous structures throughout the stellar halo of the Milky Way, based on a high-resolution chronographic (age) map, and compare with predictions of galaxy-formation models.

The formation and evolution of the recognized stellar components of the Milky Way – its central bulge, disk, and halo – are among the most fundamental and actively explored areas in contemporary astronomy. This is due, to a great extent, to the rapid expansion of photometric and spectroscopic information acquired by recent large-scale surveys such as the Sloan Digital Sky Survey[1] (SDSS), the Radial Velocity Experiment[2] (RAVE), and the Gaia-ESO Survey[3] (GES), as well as previous dedicated searches for chemically-primitive stars by the HK Survey[4,5] and Hamburg/ESO Survey[6]. The moderate- to high-resolution spectroscopy from such surveys provides the basic data, such as stellar atmospheric parameters (effective temperature, surface gravity, and metal abundance, often parametrized as [Fe/H]; 'metals' are understood to mean the chemical elements beyond H and He) and radial velocities, which can be used to derive kinematic and chemical constraints on the main structures and stellar populations of the Milky Way. What has been missing, until only recently, is the ability to assign ages to individual stellar populations, so that the full chemo-dynamical history of the Milky Way can be assessed.

In the case of the Galactic bulge, crude age information has been inferred from comparison with population-synthesis models[7], while for the disk, the burgeoning field of asteroseismology provides age estimates for individual stars based on observations of their internal oscillation modes[8]. For the halo of the Galaxy, its age (and metallicity) structure was first motivated by study of the colour-magnitude diagrams for a handful of globular clusters (compact, spherical groups of hundreds of thousands to millions of stars), and used to suggest a hierarchical assembly model of the Galaxy[9], contrasting with the rapid monolithic collapse model that was initially adopted[10]. In later studies with improved precision, it was demonstrated that the majority of globular clusters located within 15-20 kpc of the Galactic center are older (by ~2 Gyr) than most of the clusters located farther away[11-14].

In the last decade, numerous studies have revealed that the stellar halo of our Galaxy is a complex system comprising at least two diffuse components with differing spatial distributions, chemistry, and kinematics, the inner-halo and outer-halo populations, along with a number of individual over-densities and stellar debris streams[15]. The over-densities, such as those identified in the directions of Virgo and Monoceros, in addition to being spatially distinguishable from the diffuse components, exhibit distinct metallicity distributions and coherent motion. The debris streams, a number of which can be directly associated with parent dwarf galaxies that are in the process of being accreted by the Milky Way, such as the Sagittarius Stream[16] and the Orphan Stream[17], also possess distinct metallicities and kinematics. The integrated contribution from these debris streams may comprise as much as half of the stars now found in the halo system[18]. In order to understand the complete assembly process of the halo system, the ability to infer ages for the various components and structures is clearly required.

A seminal study, conducted a quarter century ago, suggested a technique that can provide information on the ages of the stellar populations in the halo of the Milky Way based on the colours of a class of stars known as field blue horizontal-branch (BHB) stars[19]. Originally identified because of their distinctive position in the colour-magnitude diagrams of globular clusters, BHB stars are less massive than the Sun (0.6-0.7 $M_{Sun}$[20]), but (owing to their larger ages) have already passed through their main-sequence (core hydrogen-burning) and giant-branch stages of evolution, and are burning helium in their cores. This first study, which included some 500 field BHB stars in the range 2-12 kpc from the Galactic center, reported a small, but statistically significant, shift in the mean B-V colours that the authors argued was associated with a difference in age of about 2.5 Gyrs, with the oldest stars found closest to the center.

More recently, the age structure of the halo system, based on a significantly larger sample (~4700) of field BHB stars with available spectroscopy from SDSS Data Release 8[21] (DR8), has been explored[22]. Spectroscopic data have the great advantage that, in addition to enabling estimates of the metallicity for stars in the sample, the derived stellar temperatures and surface gravities can be used to eliminate the primary contaminant population occupying a similar colour range as BHB stars, the blue straggler stars (BSSs; higher-gravity stars lying blueward of the halo main-sequence turnoff region). This spectroscopically-confirmed sample of BHB stars was employed to construct a low-resolution chronographic (age) map of the Galactic halo system, including stars with distances up to ~25-30 kpc from the Galactic center; the region where clear inference of age structure could be obtained was limited to ~25 kpc. In addition to taking into account modern models of BHB evolution to derive the dependence of BHB colours on age, this study was also able to demonstrate that the colour shift used to infer ages was not strongly correlated with changes in the metallicity of the populations, but is a clear age effect. In fact, the mean metallicity of BHB stars in the halo field is almost constant over the colour range adopted in our analysis (Ref. 22, top right panel of Fig. 2).

In this article we extend these techniques by employing a large sample of ~130,000 colour-selected field BHB stars from SDSS. Although spectroscopic information is not available for the great majority of these stars, we use the spectroscopic identifications of BHB stars and BSSs from the previous study[22] to specify colour ranges that minimize possible contamination from BSSs. The resulting sample is then used to construct a new, high-resolution, age map of the halo system of the Milky Way, extending to 60 kpc from the Galactic center. The refined map enables the identification of numerous new over-densities and possible debris streams, measures of the decrease in the mean age and increase in the dispersion of stellar ages with distance from the Galactic center, and opens a pathway toward examination of the age structure within individual structures, as demonstrated below.

For a fixed chemical composition, the initial colour of a BHB star is uniquely determined by its mass; a less-massive BHB star has a bluer colour. The age-colour relation for BHB stars emerges because older stars possess lower initial masses than younger stars, while the mean mass lost during the preceding giant-branch phase was approximately equal. The change in colour of BHB stars as they evolve away from the zero-age horizontal-branch does not affect our colour-age transformation, as our population-synthesis tool takes this into account when it generates the sequence of models used to calculate the mean colours of BHB stars.

The slope of the age-colour relation for BHB stars does not strongly depend on metallicity, however its zero-point does. For stars of the same initial mass, a lower metallicity star arrives on the horizontal branch with a bluer colour than a higher metallicity star; this information is not included in the present analysis. However, our mapping technique is statistical, in that we marginalize over the (unknown) metallicity of at least 10 stars in each accepted pixel in our chronographic maps. Unless the metallicities of the stars in a given pixel are strongly deviant from the expected distribution for halo stars, the derived mean colour should provide reasonable estimates of the age. This has already been demonstrated in a previous study[22], which recovers similar behaviors of the change in mean colour, hence age, with Galactocentric distance for BHB stars having spectroscopically-derived metallicities, [Fe/H] > –1.75, and those with [Fe/H] < –1.75. Note that small perturbations due to metallicity variations for individual structures may nevertheless exist.

The selection of our sample of BHB stars is described in the Supplemental Material. The dependence of inferred age on colour employed in this analysis was derived by adopting a horizontal-branch stellar population-synthesis tool that provides an estimate of the age shift as a function of the mean $(g-r)_0$ colour shift, $\Delta t_9 = -16.9 \Delta < (g-r)_0 >$ (where $\Delta t_9$ represents the age shift in Gyr), over the colour range $-0.30 < (g-r)_0 < 0.0$[22].

Figure 1 shows the colour (age) maps obtained by employing the colour-selected BHB stars and the colour-age relation discussed above. In panels a and d the data are shown in the (X,Z) plane, with (0,0) located at the Galactic center (Z is the vertical distance above or below the Galactic plane), where the squares in panel d represent the mean colours (and ages) of the BHB stars in a grid of 1 kpc square pixels. Panels a, b, and c of Fig. 1 are constructed by applying a Gaussian kernel smooth (of width 3 kpc) to the accepted pixels (those with 10 or more stars). The colour scale was selected such that most of its variation is concentrated in a region encompassing $\pm 2\sigma$ from the median $(g-r)_0$ colour of the accepted pixels. The distribution shown above the colour scale is a stripe-density plot of the mean $(g-r)_0$ colour (age) for the accepted pixels; the orange bar represents the median, and the green shaded areas indicate the $\pm 2\sigma$ regions. The zero-point of this age scale (the oldest age) was set to 13.5 Gyr, consistent with the inferred age of the oldest star with a confident age determination, the subgiant HD 140283[23-25]. From left to right, the colours represent progressively younger inferred ages. Panels b and c are the colour (age) maps in the (Y,Z) and (X,Y) planes, respectively.

A number of interesting features are immediately apparent from inspection of the (smoothed) chronographic maps shown in Fig. 1. The central regions of all three maps exhibit a clear concentration of very old stars, with ages ranging from 11.5 to 12.5 Gyr (black and dark blue colours), extending out to 10-15 kpc from the Galactic center, and including locations in the halo system close to the solar vicinity, (such as the Ancient Chronographic Sphere, ACS, of stars within 15 kpc of the Galactic center[22]). Outside this region the halo system is dominated by stellar populations with average ages of ~11 Gyrs (green colours).

It is also clear from inspection of the maps that there are numerous resolved structures present throughout the halo system, some (but not all) of which have been previously identified on the basis of their number-density contrast. Known over-densities and streams are identified by labels on the maps shown in Fig. 1; locations of these are listed in Table 1 of the Supplemental Material. Many of the recognized structures are members of the northern (leading/trailing) or southern (leading/trailing) arm of the Sagittarius Stream[26,27]. Figure 2 is a cutout of the (X,Z) plane in the region of the southern arms. According to the colour (age) coding, this portion of the stream spans a range of ages from ~9.5 Gyr to ~11 Gyr, with the youngest stars concentrated in the central regions, and the oldest stars located in the outer regions. Commensurate ages for stars in this portion of the stream were derived from recent photometric and spectroscopic observations[28]; stars in the northern arm exhibit a similar age distribution.

We also identify a blue-coloured extended region in panel a (also visible in panels b and c) of Fig.1 with the Virgo stellar over-density[29] (labelled with a V). The colour (age) coding suggests that the underlying stellar population in the Virgo over-density appears to be old, with an age on the order of 11.5-12 Gyr. Recent studies of main-sequence stars located in the Virgo over-density suggest an age of ~9 Gyr[30], which is 2-3 Gyr younger than our estimate. However, the derived median metallicity ([Fe/H] = –0.7) is also substantially higher than the values reported in other recent studies, [Fe/H] = –2.0[31] and –1.5 < [Fe/H] < –2.5[32]; there remains doubt if the same stellar population is being considered.

The yellow/orange-coloured features in panels a and b of Fig. 1 may be portions of the Styx Stream[33,34] (labelled with Styx). The Orphan Stream[35] is associated with a blue-coloured feature in panel a and b (labelled with an O). Both streams cover a range of ages of 10-11 Gyr. Panel a also shows a blue-coloured region in the Southern Galactic Hemisphere identified as a portion of the Cetus-Polar Stream[36,37] (labelled as CPS). According to the colour (age) coding, the stream is ~12 Gyr old.

A yellow-coloured clump in panel b is likely a portion of the Hercules-Aquila cloud[38] (labelled with HA); according to our colour (age) coding the structure has an average age of ~11 Gyr.

There are other clumps and/or debris streams visible in panels a-c of Fig. 1 (blue or orange-yellow colors) that could not be identified with any known structures. We defer further discussion of these to a future paper, which reports on a quantitative search for detectable structures in these maps.

Although the (X,Z) chronographic map exhibits a transition from older to younger stars as the vertical distance increases, the underlying age distribution is more complex, as can be appreciated from inspection of the age distributions shown in Figure 3. The left and right columns of panels represent the regions above and below the Galactic plane, respectively. Both distributions exhibit the relative dominance of the oldest halo stars from close to the Galactic plane up to about 10 kpc, which we associate with an ACS that extends into regions of the halo including the solar vicinity. At larger distances from the plane the number of younger stars gradually increases, and the median age, represented by the vertical orange bars, moves progressively towards younger ages. It is also clear that the dispersion in age (shown in the upper left of each panel) increases with distance from the Galactic plane. Between 3 kpc and 30 kpc, the median stellar age shifts from ~11.5 Gyr to ~11.1 Gyr, yet the oldest stars remain present at all distances. This trend suggests that as one goes outward from the Galactic center the number of younger structures progressively increases, and that these spatially-distinct systems become a significant part of the halo system at distances exceeding 30 kpc. Figure 4 summarizes the variations in the mean ages of halo stars with distance in the vertical and radial directions.

Previous studies of globular clusters in the Milky Way showed that the morphology of the colour distribution of cluster horizontal-branch (HB) stars depends on the cluster metallicity, in the sense that metal-rich clusters ([Fe/H] > –0.8) typically exhibit very red HBs, while metal-poor clusters have different HB colour distributions, even at fixed metallicity[11]. This characteristic suggested that there is a second parameter, likely the cluster's age, which influences the HB morphology. The second parameter is strongly correlated with Galactocentric distance (younger clusters being found at larger distance, beyond ~ 40 kpc), and the scatter in HB morphology increases with increasing distance[9,11]. In other words, the age dispersion of globular clusters in the Milky Way's halo increases as the distance increases, in agreement with our finding for field stars in the diffuse halo, when age is assumed as the second parameter.

Contemporary ΛCDM theories of structure formation predict that galaxies formed from the hierarchical accretion and mergers of proto-galactic systems[39]. Numerical simulations of this process, which describe the joint evolution of baryons and dark matter, predict that the oldest stellar populations are mainly concentrated in the inner regions of the resulting halos, with tightly bound orbits[40,41]. This is a natural consequence of the inside-out assembly of halos, whereby progenitor halos with a distribution of masses formed at early times and combined to assemble larger fragments, which in turn merged to form the primary halo of a galaxy. Some of the stars now found in the inner regions of galaxies were born in these proto-galactic systems (or fragments) and then became part of the main galaxy halo by merging and accretion (accreted stars), while others were born from infalling gas mainly associated with progenitor galaxies (in situ stars)[42-44].

The chemical evolution of lower-mass fragments is expected to be truncated due to either the consumption of all of the available gas, or the expulsion of gas by massive-star supernova explosions, after only a limited amount of star formation has taken place (other quenching mechanisms include tidal stripping and reionization). This quenching likely occurred before these clumps merged with the rest of the proto-galaxy, hence they primarily contribute to the oldest stellar populations in the halo system[45].

In the case of more-massive fragments, the star-formation process is expected to progress further due to the larger initial gas content and the relatively deeper potential wells, which can retain gas even in the presence of multiple generations of star formation. In such environments, star formation is expected to halt only when these fragments begin to merge, or undergo later dissipational interactions with the proto-galaxy that result in either the stripping or shock-heating of the remaining gas.

A simple scenario for the formation of stellar haloes based on two different clocks can be envisaged, in order to understand the results shown in Fig. 3 – a chemical-evolution clock that operates within each progenitor sub-halo that depends on their physical properties (gas fraction, star-formation rate) and an accretion clock that tracks the assembly process within a cosmological context. Fig. 3 could be understood with this picture, where both very old stars and younger stars populate the regions in different proportions as a function of distance from the Galactic plane.

The presence of multiple stellar populations in the halos of galaxies (often referred to as the diffuse inner- and outer-halo populations) is a general feature of current numerical simulations of galaxy formation[42-44,46,47], and places the observational results that have revealed such populations in the Milky Way[48-53], M31[54], and other galaxies[55] on firmer theoretical footing. According to these models, the inner-halo population stars might have formed with a significant contribution from more-massive sub-halos (with sustained star formation), or formed in situ in the inner region from the rapid collapse of infalling gas. In contrast, a larger fraction of stars of the outer-halo population formed in lower-mass dwarf galaxies, and were brought into the main halo through disruption and accretion, resulting in a diffuse outer halo with distinct and significantly hotter kinematics. Merger events may also produce features such as debris streams which contribute relatively younger stars. One would expect to find samples of old stars arising from both populations present in the Solar Neighbourhood.

The new detail shown in our high-resolution chronographic map clearly indicates that younger structures dominate the outer region of the halo system; these structures are expected to arise from late-term (< 5-10 Gyr) merging events. The low-density outer region of the diffuse halo system is likely not as effective in erasing the signatures of such mergers, as there was not sufficient time for them to phase-space mix with the rest of the outer-halo stars already in place. Other observational studies show that these systems are typically more metal rich than the diffuse components ([Fe/H] > –1.5), and dominate the Galactic halo system at distances exceeding 30 kpc; the presence of the diffuse components at these distances is clearly evident in the observed metallicity distributions of K giants[18], which exhibits peaks at [Fe/H] ~ –1.3 (structures), [Fe/H] ~ –1.6 (inner halo), and [Fe/H] ~ –2.3 (outer halo).

Our technique for estimation of the age distribution for stellar populations in the halo of the Milky Way can be readily extended to new photometric samples assembled by numerous contemporary and future surveys, as well as to other large galaxies in the Local Group such as Andromeda and the Magellanic Clouds. Refinement of the technique to include a full grid of the variation in observed colours of BHB stars with metallicity, as well as age, is currently underway, and should prove useful for detailed analysis of the age structure of individual over-densities, debris streams, and dwarf galaxy satellites of the Milky Way.

**Acknowledgments**

The data used in this paper are drawn from the publicly available SDSS-DR8 (https://www.sdss3.org/dr8/). D.C., T.C.B., V.M.P., and G.L. acknowledge partial support for this work from grant PHY 14-30152; Physics Frontier Center/JINA Center for the Evolution of the Elements (JINA-CEE), awarded by the US National Science Foundation. Y.S.L. acknowledges support provided by the National Research Foundation of Korea to the Center for Galaxy Evolution Research (No. 2010-0027910) and partial support from the Basic Science Research Program through the National Research Foundation of Korea (NRF) funded by the Ministry of Science, ICT & Future Planning (NRF-2015R1C1A1A02036658). R.M.S. and S.R. acknowledge CAPES (PROEX), CNPq, PRPG/USP, FAPESP and INCT-A funding. P.D. acknowledges partial funding from a Natural Sciences and Engineering Research Council of Canada grant to Don VandenBerg. P.B.T. acknowledges partial support from PICT-959-2011, Fondecyt-113350, and MUN-UNAB projects.

**Author contributions**

D.C., T.C.B., V.M.P, R.M.S., G.L., and Y.S.L. performed the analysis and interpretations of the observations. The chronographic maps were assembled based on graphical techniques developed by V.M.P. P.D. carried out modelling of the mapping of BHB colours to age estimates. D.C., T.C.B., P.B.T. and J.T. carried out comparisons of the results with expectations from numerical simulations of galaxy formation. All authors discussed the results and commented on the manuscript.

**Additional information**

Supplemental information is available in the online version of the paper. Reprints and permissions information is available online at www.nature.com/reprints. Correspondence and requests for materials should be addressed to D.C.

**Competing financial interests**

The authors declare no competing financial interests.

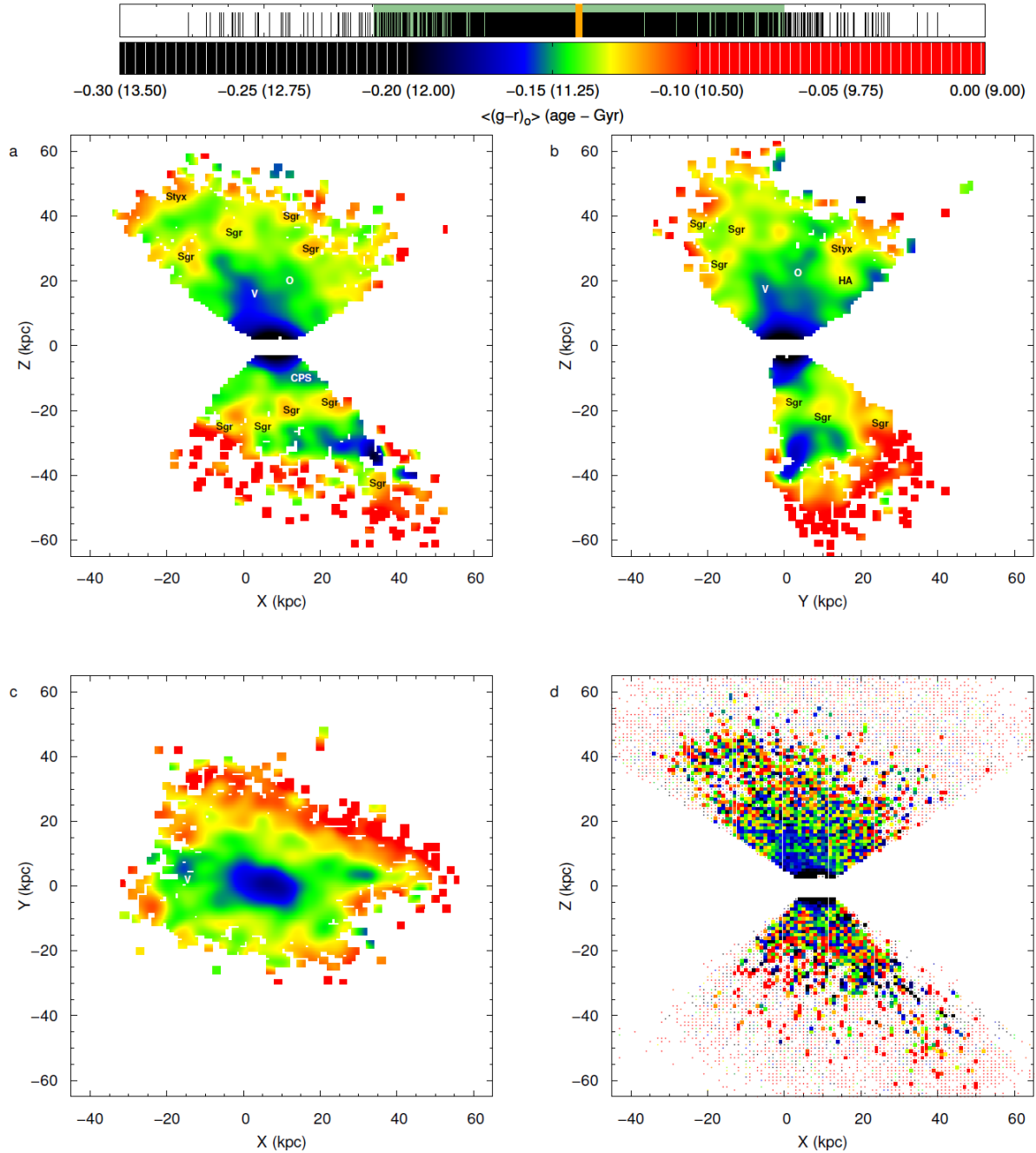

**Figure 1 | Chronographic map for photometrically-selected BHB stars from the SDSS.** The adopted Cartesian reference system is left-handed, having positive X towards the anti-center of the Galaxy. The location at (0,0,0) kpc corresponds to the center of the Galaxy, and the Sun is located at (X,Z) = (8.5,0,0) kpc. **a**, The distribution of stars in the (X,Z) plane in a square grid of 1 kpc pixels smoothed with a Gaussian kernel (of width 3 kpc) applied to the accepted pixels. The colour (age) scale is shown at the top of the diagram. The numbers in parentheses are the corresponding ages in Gyr. The orange bar is the median value of the pixels; the green-shaded area indicates the +/− 2σ dispersion region. **b,c**, Colour (age) maps in the (Y,Z) plane and (X,Y) plane, respectively. Previously known structures and over-densities are marked in panels **a-c** with black or white characters with the following meaning: Sgr: Sagittarius Stream, V: Virgo over-density, O: Orphan Stream, Styx: Styx Stream, CPS: Cetus Polar Stream, HA: Hercules-Aquila cloud. **d**, The distribution of stars in the (X,Z) plane in a square grid of 1 kpc pixels. The filled squares indicate pixels with at least 10 BHB stars (accepted pixels); the colour represents the mean $(g − r)_0$ colour for each pixel. The filled dots are pixels with less than 10 stars. The shape of the distribution reflects the selection of stars in Galactic latitude ($|b| \geq 35^\circ$).

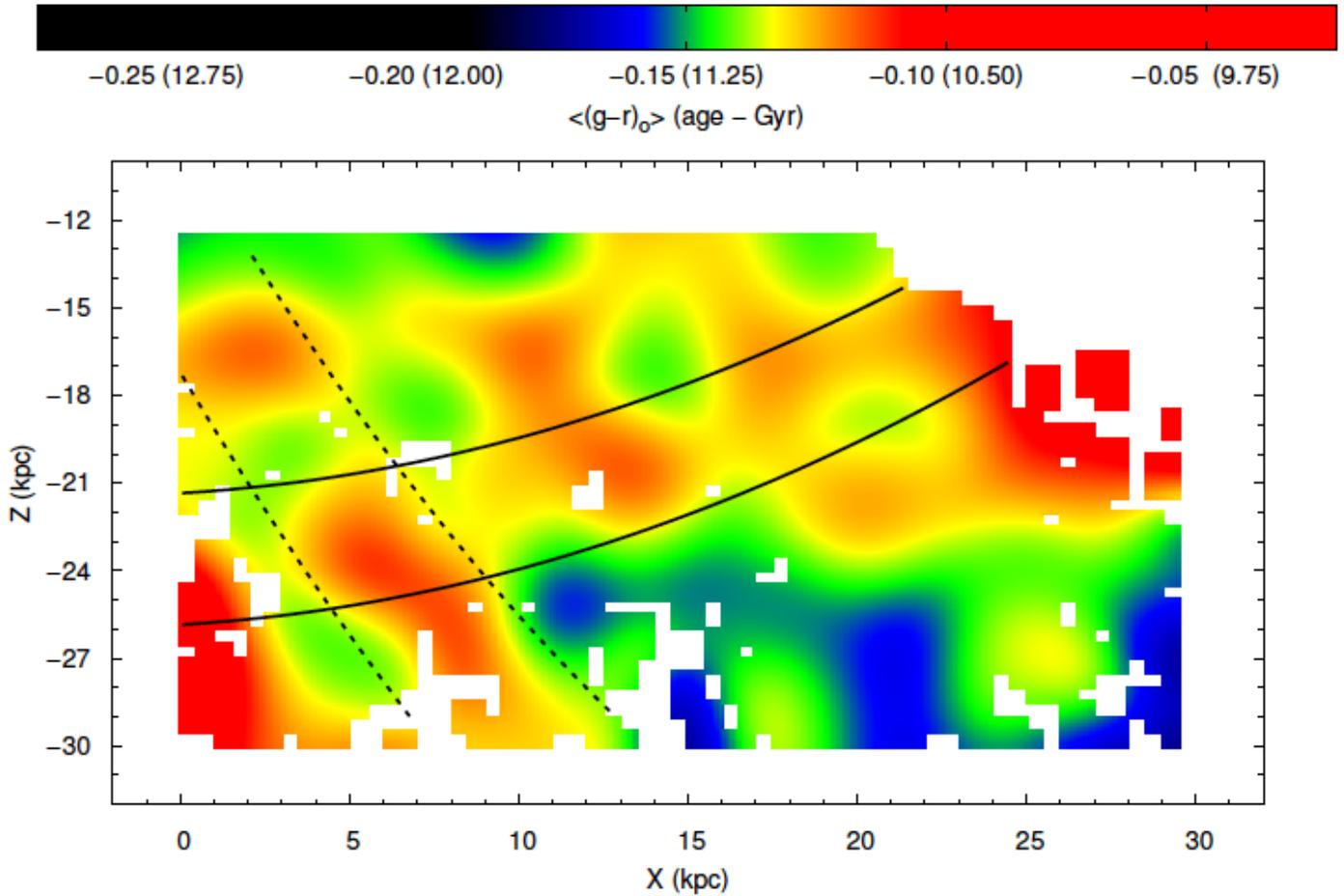

**Figure 2 | Chronographic map in the region of the southern (leading/trailing) arm of the Sagittarius Stream.** The area delimited by the black continuous curves shows roughly the location of the primary trailing debris stream, while the dashed curves delimit the location of the primary leading arm. The colour scale is the same as in Figure 1.

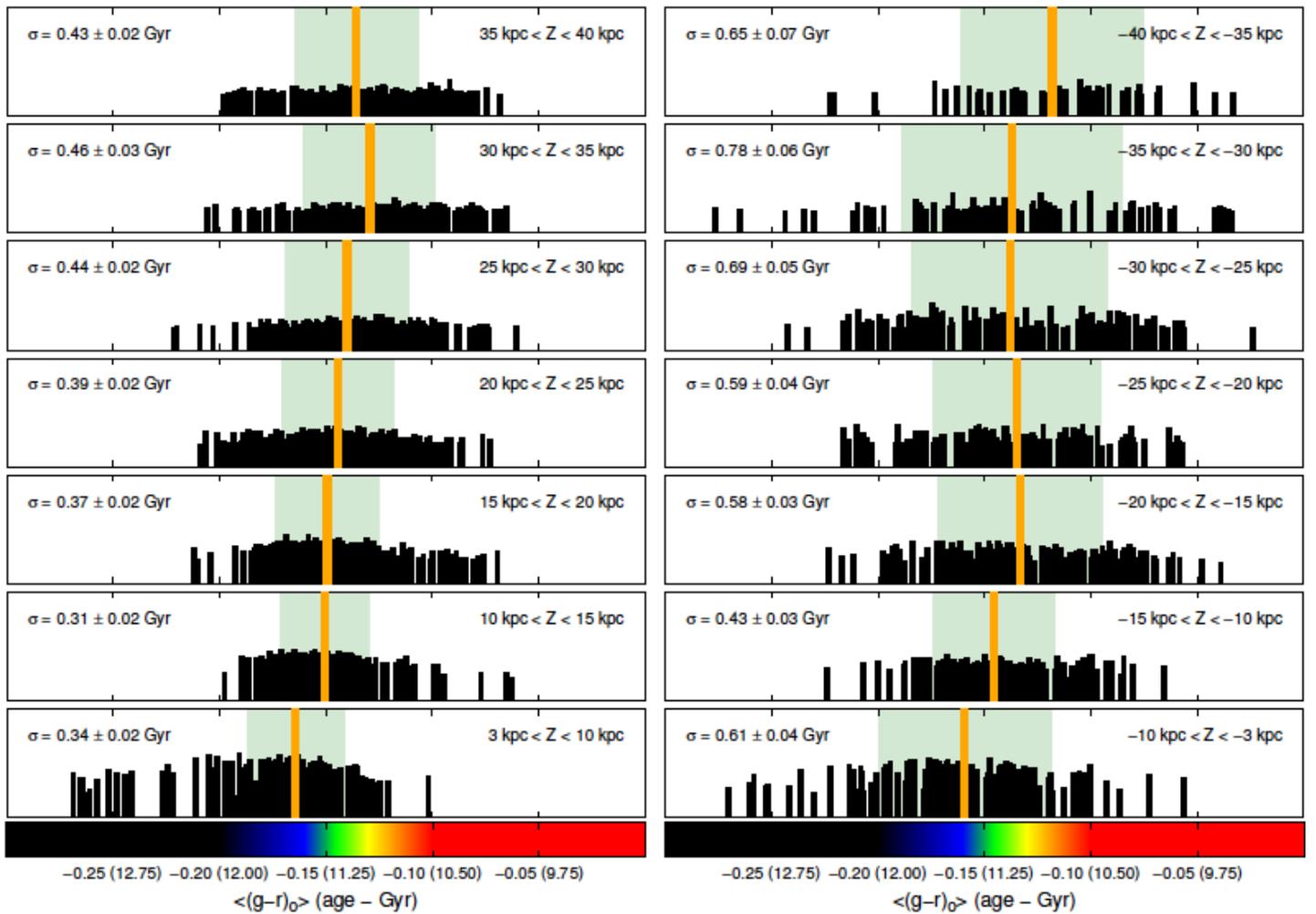

**Figure 3 | Colour (age) distribution of accepted pixels (containing at least 10 stars) for different ranges of the vertical distance, Z.** The solid black vertical bars are proportional to the logarithm of the number of stars in each pixel. The solid orange vertical bar represents the median colour (age) for each Z range. The green shaded areas encompass ±1σ units of the bi-weight estimator of scale, a robust calculation of the dispersion of the data. The colour (age) dispersions and their errors are shown in the top left of each panel. Such dispersions appear to increase with distance from the Galactic plane. The shift in median colour (age) in the Southern Galactic Hemisphere for −35 kpc < Z < −25 kpc is due to the prominent strip of older stars seen in Fig. 1. This trend is not evident in the Northern Galactic Hemisphere, and suggests an asymmetric age distribution between the two hemispheres.

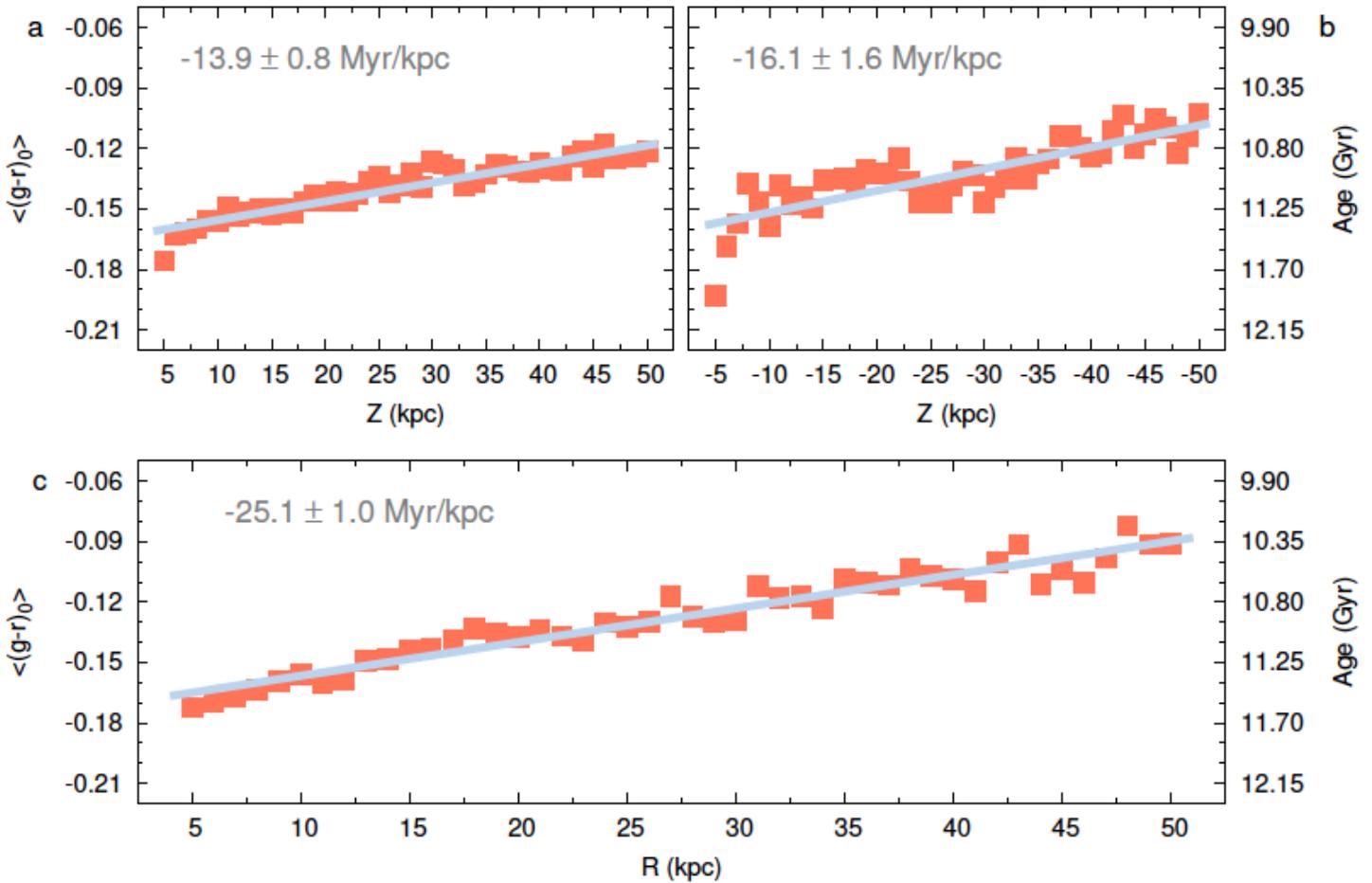

**Figure 4 | Colour (age) gradient in the vertical distance and in the radial distance directions.** The red squares denote the mean colour for each 1 kpc bin in distance (left Y axes) and the grey line represents a linear regression to the data. The right Y axes shows the mean age. **a**, Colour (age) gradient and its error above the Galactic plane. **b**, Colour (age) gradient and its error below the Galactic plane. The data indicate that the underlying stellar populations become progressively younger with distance from the Galactic plane, with a slope of −13.9 ± 0.8 Myr/kpc in the Northern Galactic Hemisphere and −16.1 ± 1.6 Myr/kpc in the Southern Galactic Hemisphere, respectively. The mean values in the range of −35 kpc < Z < −25 kpc are slightly displaced from the regression line, as discussed in Fig. 3. **c**, Colour (age) gradient over radial Galactocentric distance, R, with slope −25.1 ± 1.0 Myr/kpc.

**Supplemental material**

Candidate BHB stars were selected from SDSS-DR8 based on a first-pass colour window designed to exclude cooler (primarily main-sequence and main-sequence turnoff) stars[56]: $0.60 < (u - g)_0 < 1.60$ and $-0.50 < (g - r)_0 < 0.05$, where u, g, and r are three of the five filters used by the SDSS (u,g,r,i,z, centered at 354 nm, 477 nm, 623 nm, 762 nm and 913 nm, respectively). Note that all magnitudes and colours used in this work have been corrected for interstellar absorption and reddening[57,58]; stars with reddening estimates in excess of 0.1 mag have been excluded. We also selected stars with Galactic latitude $|b| \geq 35°$ in order to reduce contamination from the disk system of the Galaxy. BHB stars are relatively luminous, and their intrinsic (absolute) magnitudes do not strongly vary as a function of [Fe/H][56]. These absolute magnitudes, required in order to derive estimates of the BHB distances, were obtained as a function of their $(g - r)_0$ colours using a fourth-order polynomial[59].

We have also considered possible contamination from white dwarfs (WDs) and quasars (QSOs). In the colour-colour diagram $(u - g)_0$ vs. $(g - r)_0$, WDs are typically located in the ranges $-0.6 < (u - g)_0 < 0.6$ and $-0.5 < (g - r)_0 < 0.3$[60], which lies outside our selection area. They are also intrinsically much fainter than BHB stars; they are unlikely to be present in our sample, given that it only comprises stars with magnitudes $u_0 < 20$ and $g_0 < 20$.

In the case of QSOs, the colour region they occupy depends on their redshift[61]. Low-redshift ($z < 2.2$) QSOs have $(u - g)_0 < 0.6$ and mid-redshift ($2.2 < z < 3$) QSOs have $0.6 < (u - g)_0 < 1.5$ and $0.0 < (g - r)_0 < 0.2$. These intervals are outside the colour region adopted for our BHB selection. High-redshift ($z > 3.0$) QSOs are generally faint, with $u_0 > 20.6$ and $(u - g)_0 > 1.5$; at $z > 4.5$, the corresponding magnitudes are $u_0 > 21.5$ and $g_0 > 21.0$. These magnitudes are fainter than our adopted limiting magnitude, and their $(u - g)_0$ values are outside the range employed in our analysis. We conclude that it is very unlikely that QSOs are included in our BHB selection.

Observational errors in the colours play an important role in the selection because the $(u - g)_0$ colour is used to separate BHB stars from BSSs, and the transformation between the colours of BHB stars and their ages is derived using $(g - r)_0$. For stars in our sample with g- and r-band apparent magnitudes $g_0, r_0 < 20$, the majority have errors in $(g - r)_0$ colour well below 0.1 mag (only 4% of the stars possess errors > 0.1); the average error is ~ 0.017 mag. The average error in $(u - g)_0$ colour is on the order of ~ 0.04 mag (the larger error in $(u - g)_0$ colour is due primarily to greater uncertainties on the u-band magnitude). We only include stars with errors in $(g - r)_0$ and $(u - g)_0$ colours less than 0.10 mag and 0.15 mag, respectively. At fainter apparent magnitudes ($g_0, r_0 > 20$) the average errors in the colours increases, to on the order of ~0.1 mag and ~0.2 mag in $(g - r)_0$ and $(u - g)_0$, respectively. For this analysis we have only included stars with magnitudes $g_0, r_0 < 20$ in order to reduce the uncertainties on colours.

The adopted $(g - r)_0$ colour window for our selection of the BHB stars is $-0.30 < (g - r)_0 < 0.0$, which was shown to increase the dynamical range of the observed colour shift compared to the narrower window that was adopted in previous works[62,63].

Supplemental Figure 1 shows the colour-colour diagram, $(u - g)_0$ vs. $(g - r)_0$, for the spectroscopically-confirmed BHB stars and BSSs (panel a), and for the sample of candidate BHB stars and BSSs initially selected from the SDSS photometric database (panel b). From inspection of panel a, it is clear that the samples of stars classified as BHB stars and BSSs exhibit a characteristic concave shape, but they overlap in the colour-colour diagram (in particular for stars with colours $(g - r)_0 > -0.15$). Locally weighted regression fits were applied to both samples in order to derive the average location of BHB stars and BSSs in this diagram.

The number of stars satisfying the initial colour cuts is 901,627, which is reduced to ~562,000 stars after application of the Galactic latitude and colour-error cuts (panel b of Supplemental Figure 1). The most prominent feature in this panel is the concave shape comprising the overlapping candidate BHB stars and BSSs.

The red and cyan curves are the average locations of the spectroscopically-confirmed BSSs and BHB stars shown in the top panel; the yellow line is the adopted lower limit for classification as a BHB star. The spectroscopic sample was employed to evaluate the level of contamination from BSSs as a function of $(g-r)_0$ colour. In the colour range $-0.2 < (g-r)_0 < 0.0$ the contamination is less than 15% for stars located beneath the BHB local regression line. For stars with bluer colours, $(g-r)_0 < -0.2$, the level of contamination from BSSs decreases, and BHB stars can be chosen above the average location represented by the regression line. The cyan dots indicate the upper limits on the $(u-g)_0$ colour in selected narrow intervals of $(g-r)_0$, below which the BSS contamination is less than 15%. After application of these cuts, the final selection of likely field BHB stars for our remaining analysis includes N ~130,000 stars.

**Identification of structures and over-densities in the color (age) map**

The recognized over-densities and debris streams labelled in Figure 1 of the main material are listed in Table 1. The identification is based on the location of the various features emerging in the color (age) map in the (X,Y,Z) Cartesian reference frame with (0,0,0) in the center of the Galaxy. Some features associated with the leading and trailing arms of the Sagittarius Stream are based on the location predicted by models[64].

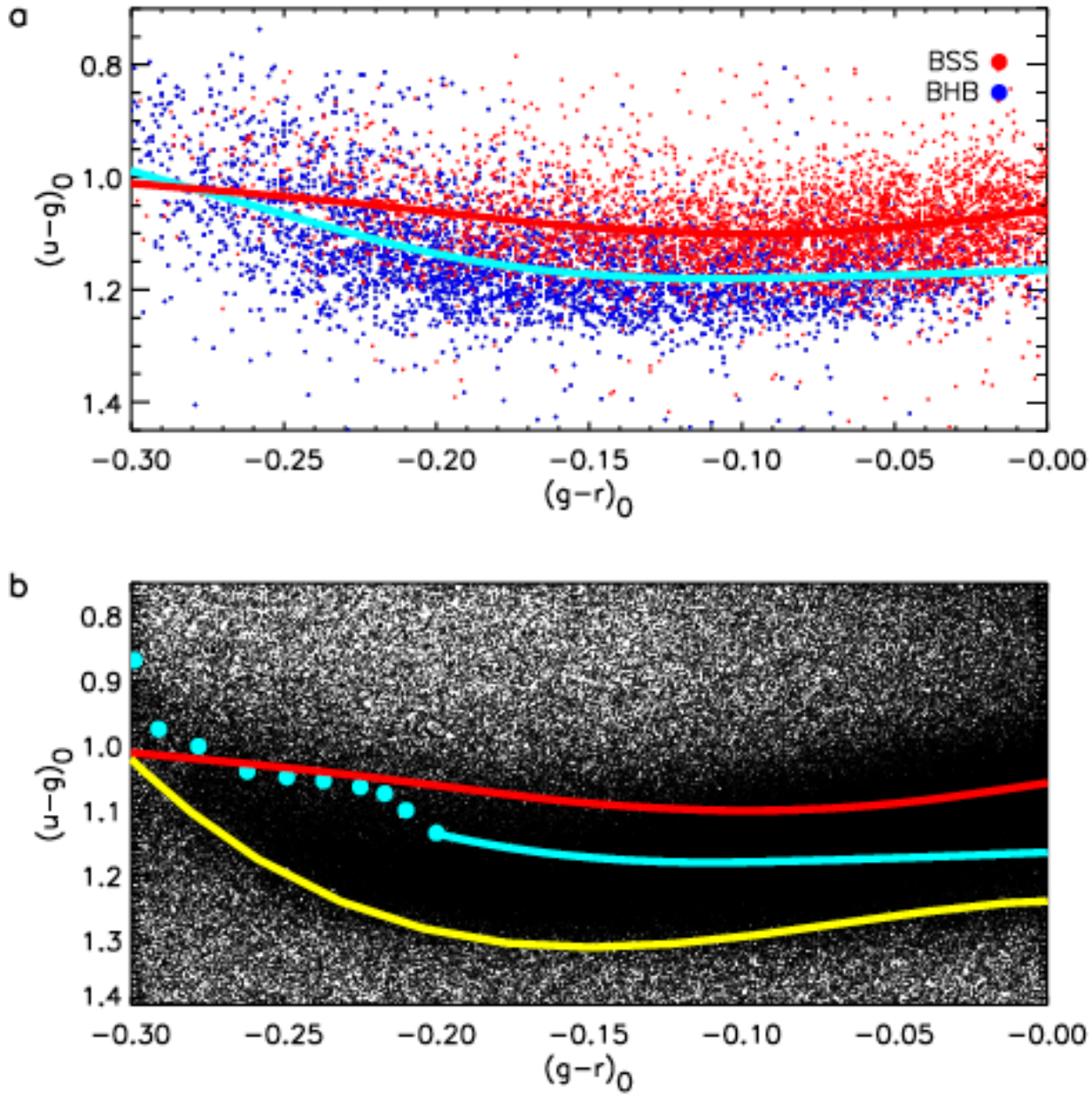

**Supplemental Figure 1 | Colour-colour diagrams used for selection of BHB stars in our sample. a,** Spectroscopically-confirmed BSSs and BHB stars are indicated with red and blue dots, respectively[48]. The two curves show local regression fits to the samples, and represent a useful indicator of the average location of the two stellar types. **b**, The sample of stars selected from SDSS-DR8 (black dots) that pass the specified cuts in colour, Galactic latitude, and colour errors (see text). The red and cyan curves are the same as in panel a. The lower yellow curve indicates the boundary of the region where the BHB stars reside. The cyan dots indicate the upper limit of $(u-g)_0$ for each $(g-r)_0$ colour interval for bluer stars in this region, where the selection of BHB stars can be carried out with < 15% contamination from BSSs.

**Table 1.** Location of the recognized structures in the Galactocentric Cartesian reference frame

| Structure/Over-density | Plane (X, Z) (kpc) | Plane (Y,Z) (kpc) | Plane (X,Y) (kpc) |
|---|---|---|---|
| Sagittarius North (Leading/Trailing Arm) | (-12,27); (4,35) (12;37); (18,30) | (-16,37); (-12,23) (-10,35) | |
| Sagittarius South (Leading/Trailing Arms) | (0,-25);(6,-25); (12,-20) (23,-18); (35,-45) | (0,-18);(10,-22) (20,-25) | |
| Virgo Over-density | (5,16) | (-5,16) | (-15,5) |
| Styx Stream | (-12,45) | (15,30) | |
| Orphan Stream | (13,18) | (6,23) | |
| Hercules-Aquila Cloud | | (12,22) | |
| Cetus-Polar Stream | (15,-11) | | |